\documentclass[a4paper,12pt,oneside]{article}
\usepackage{amsmath}
\usepackage{amsthm}
\usepackage{amscd}
\usepackage{color}
\usepackage{amssymb}
\usepackage[utf8]{inputenc}
\usepackage[english]{babel}
\usepackage{graphicx}
\usepackage{epsfig}
\usepackage{fancyhdr}
\usepackage{authblk}
\usepackage{blindtext}
\usepackage{typearea}
\usepackage{longtable}
\def\be{\begin{equation}}
\def\ee{\end{equation}}
\def\es{\end{section}}
\def\bc{\begin{center}}
\def\ec{\end{center}}
\def\bp{\begin{pro}}
\def\ep{\end{pro}}
\def\p{\partial}
\def\n{$n-$}
\def\e{\eta}
\def\x{\xi}
\def\pi{\phi}
\def\ov{\overline}
\def\al{\alpha}
\def\blm{\begin{lm}}
\def\elm{\end{lm}}
\def\bpmat{\begin{pmatrix}}
\def\epmat{\end{pmatrix}}
\def\bvmat{\begin{vmatrix}}
\def\evmat{\end{vmatrix}}
\def\maf{\mathfrak}

\newtheorem{defi}{Definition}[section]
\newtheorem{exm}{Example}[section]

\newtheorem{lm}{Lemma}[section]

\newtheorem{pro}{Proposition}[section]
\numberwithin{equation}{section}

\begin{document}
\begin{center}
{\bf\Large Symmetry Classification of Scalar $n$th Order  Ordinary Differential Equations}\\
\null\vspace*{0.6cm}
Said Waqas Shah$^1$,  F. M. Mahomed$^{2}$ and
H. Azad$^1$\\[3ex]
$^{1}$Abdus Salam School Of Mathematical Sciences GC University, Lahore. 68-B, New Muslim Town, Lahore 54600, Pakistan\\
Emails: waqas.shah@sms.edu.pk\quad Hasan.Azad@sms.edu.pk \\[2ex]
$^2$DSI-NRF Centre of Excellence in Mathematical and Statistical Sciences,
 School of Computer Science
and Applied Mathematics, University of the Witwatersrand,
Johannesburg, Wits 2050, South Africa\\
Email: Fazal.Mahomed@wits.ac.za

\end{center}

\textbf{Abstract:}  We complete the Lie symmetry classification of scalar $n$th order, $n\ge4$, ordinary differential equations by means of the symmetry Lie algebras they admit. It is known that there are three types of such equations depending upon the symmetry algebra they possess, viz. first-order equations which admit infinite dimensional Lie algebra of point symmetries, second-order equations possessing the maximum eight point symmetries and higher-order, $n\ge3$, admitting the maximum $n+4$ dimensional symmetry algebra. We show that $n$th order equations for $n\ge4$ do not admit maximally an $n+3$ dimensional  Lie algebra except for $n=5$ which can admit $sl(3,R)$ algebra. Also, they can possess an $n+2$ dimensional Lie algebra that gives rise to a nonlinear equation that is not linearizable via a point transformation. It is shown that for $n\geq5$ there is only one such class of equations.\\


\begin{section}{Introduction}

Symmetry Lie algebras of scalar $n$th order ordinary differential equations (ODEs) have been extensively studied over several years since the initial ground breaking works of Lie \cite{Lie1,Lie2}. Equations of order one are equivalent to each other via point transformation. For scalar higher order ODEs, Lie \cite{Lie1} proved that the maximum dimension of the point symmetry Lie algebra for a scalar second order ODE is eight dimensional and occurs for linear and linearizable by point transformation equations. Lie \cite{Lie2} obtained a complex classification of second order ODEs in terms of their Lie algebras. Mahomed and Leach \cite{Mah} derived the real classification and showed that a second order equation can admit 0, 1, 2, 3 or the maximum 8 dimension real symmetry algebra. The original Lie classification and the classification in the real domain \cite{Mah}  for second order ODEs are, inter alia, compared in \cite{Ibr}. Algebraic linearizability criteria were initiated by Lie \cite{Lie1},  who showed that such second order equations possessing a Lie algebra of dimension 2 and of rank one, are linearizable. This falls under Type II and Type IV canonical forms in Lie’s classification. The Types I and III cases with focus on linearizability were achieved in \cite{Sar}, \cite{Lea}. The reader is also referred to the survey article \cite{Mah1}. We furthermore deduce all the canonical forms for $n$th order, $n\ge4$, equations that admit $n$ dimensional symmetry algebras. The $n=2,3$ cases are well-known in the literature. \\

In the study of scalar linear ODEs of order $n$, $n \ge 3$, Mahomed and Leach \cite{Mah2} (see also \cite{Ibr}, \cite{Mah1} as well as the contribution by Krause and Michel \cite{Kra}) demonstrated that the point symmetry algebra can be $n+1, n+ 2$ or $n+4$. Thus, for $n \ge 3$, scalar linear ODEs are not necessarily equivalent to each other via point transformation. Moreover, for $n \ge 3$, there exist nonlinear as well as nonlinearizable ODEs with $n+2$ and $n+3$ symmetry algebras \cite{Mah2}. It is important to remark that second order ODEs are quite different to higher order equations $n \ge 3$ as per the symmetry algebras they admit.  Apart, from the maximum dimension of the point symmetry algebra being 2+6 for second order ODEs and that for higher order, $n \ge 3$, equations $n + 4$ (see Lie \cite{Lie1,Lie2}), there are two more notable differences. Secondly, all linear second order ODEs are equivalent to the free particle equation whereas a linear higher order $n \ge  3$ ODE has been determined to have three equivalence classes depending upon whether it has $n + 1, n + 2$ or the maximum number $n + 4$ of point symmetries \cite{Mah2}. Thirdly, the complete or full algebra of symmetries of a second order ODE is a subalgebra of its maximum algebra $s1(3, R)$, whereas the full algebra of a higher order $n \ge 3$ ODE is not necessarily a subalgebra of its maximum Lie algebra \cite{Mah2}.\\

The symmetry Lie algebra classification of third order ODEs  as well as linearization by point transformation have been investigated in a number of relevant publications (see \cite{Mah2,Kra,Che,Mah3,Gat,Gre,Mel,Neu,Dwe1,Dwe2,Dwe3}). Furthermore, integrability and reductions for third order ODEs were looked at in \cite{Nuc1}.\\

Scalar fourth order ODEs were considered in recent works from the point of view of Lie classification in terms of four-dimensional algebras, canonical forms as well as integrability (see \cite{Nuc2,Ayu,Waq1}. A complete Lie classification and algebraic linearization were also attempted in \cite{SFHT}. Linearization criteria, by point transformation, for such ODEs were found in \cite{Suk}.\\

This work is a continuation of our previous work on the classification of scalar fourth order ODEs \cite{SFHT}. Here we go on to complete the classification of $n$th order ODEs according to the symmetry algebras they admit. We emphasize that we use the same tools and scheme as we have for scalar fourth order ODEs but in general form as the order of the equations considered here are for any $n > 4$.\\


The first section deals with notation which we have taken from \cite{SFHT} as well as the established methods used in this paper. In the second and third sections we classify all equations of order $n$ which admit $(n+1)$- and $(n+2)$-dimensional Lie algebras, respectively. Thereafter we find the $(n-1)$- and $n$-dimensional fundamental invariants of algebras of dimension $n$. In the last section we provide some applications and present concluding remarks.

\section{Notation and Methods}

By $(m, n)$ we denote the type of algebra where $m$ is the type of algebra in \cite{GKO} and $n$ is the dimension of that algebra. We denote a general vector field or generator as
\begin{center}
  $X_{i} = \xi_{i}(x, y)\partial_{x} + \eta_{i}(x, y)\partial_{y}$, \hspace{1cm} $i = 1,2,...,n$.
\end{center}
Here $(x, y) \in \mathbb{R}^{2}$ and $\partial_{x}$ denotes $\frac{\partial}{\partial x}$. Note also that $n$ is the dimension of a Lie algebra of which $X_{i}$s are the generators.
Other notation will be mentioned as they arise.

\subsection{Classification Scheme}

Let $L$ be an $m$-dimensional Lie subalgebra of vector fields. To find an invariant equation of order $n$ we consider the normal form of an $n$th order ODE as
\be
y^{(n)} = H(x, y, y',y'',..., y^{(n-1)}).
\ee \label{0.1}\\
For a generator $X$ to be a symmetry of \eqref{0.1} the following condition must hold:
\be
pr_{n}(X)(y^{(n)}-H)|_{y^{(n)} = H} = 0
\ee \label{0.2}\\
where $pr_{n}$ denotes the $n$th prolongation and $|_{y^{(n)} = H}$ denotes the constraint. In this case the resulting $n$th order equation has the most general form of an $n$th order equation invariant under $X$. If the condition \eqref{0.2} holds for every $X_{i}$ , $i = 1,2,..., m$, then the resulting $n$th order equation is called the invariant equation of $L$ and $L$ is called the symmetry algebra of such an equation. This will be more elaborated in the next sections.

\subsection{Fundamental Invariants}

Let $L$ be an $m$-dimensional Lie subalgebra of vector fields defined on a subspace $\mathbb{D} \subset \mathbb{R}^{2}$. Then the $n$th order prolonged Lie algebra is defined on a subspace $\mathbb{D}^{(n)} \subset \mathbb{R}^{n+2}$. Suppose that $r_{0}$ is the rank of $L$ in $\mathbb{D}$. Then $r_{n}$ will be the rank of the prolonged $L$ in $\mathbb{D}^{(n)}$. The rank here means the rank of the matrix whose rows are coefficients of the $m$ generating vector fields of $L$. \\

 Now let $d_{n}$ be the number of differential invariants of order $n$.  Then we have
 \begin{center}
 $d_{n} = n + 2 - r_{n}$,  \ \  $n\geq 0$
 \end{center}
 \begin{exm}
 Consider $X = \partial_{x}$. Here $n =0$ and $r_{o} = 1$. Thus $d_{0} = 0 + 2 - 1 = 1$, so we have one zeroth order differential invariant which is $u = y$.
 \end{exm}
 \begin{exm}
 Consider $X_{1} = \partial_{x}$, $X_{2} = \partial_{y}$. Again $n =0$ , $r_{o} = 2$ and $d_{0} = 0 + 2 - 2 = 0$,  and there is no zeroth order invariant. Now for the first prolongations of $X_{1}$ and $X_{2}$,  $n = 1$, $r_{1} = 2$ and $d_{1} = 1 +2 -2 = 1$. Thus there is a first order differential invariant $ u = y'$.
 \end{exm}
A very important work for finding the fundamental differential invariants for \n th order equations is done by Neseterenko in \cite{MN}.
 \subsection{Invariant Differentiation Operator}

 If $u$ , $v$ are invariants then by Lie's theorem, $\frac{D_{x}v}{D_{x}u}$ is also an invariant. This process is also called invariant differentiation \cite{IBV}. Recall that $D_{x}$ is the total differentiation operator. We can write this as
 \begin{center}
 $\dfrac{D_{x}v}{D_{x}u} = (D_{x}u)^{-1}D_{x}v$\\
                        = $\lambda D_{x}v$
 \end{center}
 We call $\lambda D_{x} = \textit{D}$ the invariant differentiation operator once we know $\lambda$. \\
 \newline
 Suppose we have an unknown $\lambda(x, y, y', ..., y^{(n)})$. We require $\lambda D_{x}v$ to be invariant, and therefore we need
 \begin{equation}\label{2.3}
 pr_{n}(X)(\lambda D_{x}v) = 0
 \end{equation}
 For the time being we ignore the prolongation symbol and consider $pr_{n}(X)$ simply as $X$. We can write
 \begin{center}
 $X = \overline{X} + \xi D_{x}$
 \end{center}
 where $\overline{X} = w\partial_{y} + D_{x}w \partial_{y'} + D_{x}^{2}w \partial_{y''} + ... + D_{x}^{n}w \partial_{y^{(n)}}$ is called the canonical operator and  $w = \eta - \xi D_{x}$. The equation \eqref{2.3} is then written as
 \begin{center}
 $(\overline{X}+\xi D_{x})(\lambda D_{x}v) = 0$\\
 $\Rightarrow \xi D_{x}(\lambda D_{x}v) + \overline{X}(\lambda)D_{x}v + \lambda \overline{X}(D_{x}v) = 0$\\
 $\Rightarrow \xi D_{x}\lambda D_{x}v + \xi \lambda D_{x}^{2}v + (X\lambda - \xi D_{x}\lambda)D_{x}v + \lambda \overline{X}(D_{x}v) = 0$\\
 $\Rightarrow \xi D_{x} \lambda D_{x}v + \lambda \xi D_{x}^{2}v + X(\lambda) D_{x}v - \xi D_{x}\lambda D_{x}v + \lambda \overline{X}(D_{x}v) = 0$\\
 $\Rightarrow \xi \lambda D_{x}^{2}v + X(\lambda)D_{x}v + \lambda D_{x}(Xv - \xi D_{x}v) = 0$\\
 $ \Rightarrow (X(\lambda) - \lambda D_{x}\xi)D_{x}v = 0$\\
\end{center}

 This gives:
 \begin{equation}\label{*}
 X(\lambda) = \lambda D_{x} \xi.
 \end{equation}
 Hence, $\lambda$ satisfies the non-homogenous linear PDE \eqref{*}. One only requires one nontrivial solution for $\lambda$ which can be a constant as well.
 \begin{exm}
 Let $X_{1} = \partial_{x}$ and $X_{2} = \partial_{y}$. We know that $u = y'$ is a first order differential invariant of $X_{1}$ and $X_{2}$. Applying the condition \eqref{*} we have
 \begin{center}
 $X_{1}\lambda = 0$ , $X_{2} \lambda = 0$
 \end{center}
 This clearly shows that we can set $\lambda = 1$ and therefore the invariant differentiation operator can be taken as $\textit{D} = (1)D_{x} = D_{x}$.
Thus
\begin{center}
$\textit{D}y' = D_{x}y' = y''$
\end{center}
is a second order differential invariant.
\end{exm}
\subsection{Lie Determinants}

\begin{defi}
Consider an $m$-dimensional Lie subalgebra of vector fields whose generators are given as
\begin{center}
$X_{k} = \xi_{k}\partial_{x} + \eta_{k} \partial_{y}$,
$k = 1, 2, ... , m$.
\end{center}
Then the determinant of the following matrix
\bc
$\bpmat
\xi_{1} & \eta_{1} & \eta^{[1]}_{1} & \eta^{[2]}_{1} &\cdots & \eta^{[m-2]}_{1} \\
\xi_{2} & \eta_{2} & \eta^{[1]}_{2} & \eta^{[2]}_{2} &\cdots & \eta^{[m-2]}_{2}\\
\vdots  & \vdots  & \ddots & \vdots  \\
\xi_{m} & \eta_{m} & \eta^{[1]}_{m} & \eta^{[2]}_{m} &\cdots & \eta^{[m-2]}_{m}
\epmat$
\ec
is called the Lie determinant corresponds to the $m$-dimensional Lie algebra for $m\geq 2$.
We denote the Lie determinant by $\Lambda_{L}$.
\end{defi}

Lie proved that for an $m$-dimensional Lie subalgebra of vector fields $L$, the Lie determinant gives rise to all the invariant equations of order $\leq m - 2$ \cite{LIE2}. Similarly it can be noticed that the rank of the prolonged algebra is $m$, i.e. maximal unless the Lie determinant vanishes in which case the rank of $L$ diminishes. Here the algebra is prolonged up to order $m-2$. These invariant equations are called the singular invariant equations of $L$. The fundamental differential invariants which are not singular must be of order $m-1$ and $m$ and the higher order differential invariants can be then be found from the fundamental invariants.\\
\newline
Note that here, once we find the $(m-1)$th order differential invariant say $u$ and the invariant differentiation operator $\textit{D} = \lambda D_{x}$, we can determine the $m$th order differential invariant as it is none other than $\textit{D}u$.

\section{(n+1) Dimensional Algebras}
\subsection{Nonlinear Equations}

It is easy to observe that $(20, n+1)$ is not admitted by an equation as there can only be $n$ solution symmetries. The linearizable case $(21,n+1)$ is studied in the next subsection 3.2. Also, $(22,n+1)$ is not possessed as a maximal Lie algebra. We consider the other $n+1$ dimensional algebras.\\

$(24, n+1), r = n-2$, $n\geq3$: $X_{1} = \p_{x}$, $X_{2} = \p_{y}$, $X_{3} = x \p_{x} + \alpha y \p_{y}$, $X_{4} = x \p_{y}$, $X_{5} = x^{2} \p_{y}$, ..., $X_{n+1} = x^{n-2} \p_{y}$.\\

The generators, except $X_{3}$, imply that a general $n$th order equation of the type \eqref{0.1} admitting these generators must be of the form:
\be
y^{(n)} = H(y^{(n-1)}).\label{1.1}
\ee
The $n$th prolongation of $X_{3}$ is: $x \p_{x} + \alpha y \p_{y} +(\alpha - 1)y' \p_{y'} + ,...,+(\alpha - n)y^{(n)} \p_{y^{(n)}}$.\\

Applying this to the equation \eqref{1.1} and solving we find the general form of an $n$th order equation admitting this algebra to be
\be
y^{(n)} = K ({y^{(n-1)}})^ {\alpha - n \over \alpha - n + 1},\,K\ne0,\label{1.2}
\ee
where $K\ne0$ is an arbitrary constant and $\alpha \neq n-1$. For $K=0$ or
$\alpha = n-1$, it is easy to see that the general form of such an equation admitting this algebra must be $y^{(n)} = 0$, though we know that such an equation admits the maximal
$(n + 4)$-dimensional algebra of which such an algebra is the subalgebra.\\

$(25, n+1), r = n-1$, $n\geq 2$: $X_{1} = \p_{x}$, $X_{2} = \p_{y}$, $X_{3} = x \p_{y}$, $X_{4} = x^{2}\p_{y}$ , ..., $X_{n} = x^{n-2} \p_{y}$, $X_{n+1} = x \p_{x} + (ry + x^{r})\p_{y}$.\\

The $n$th prolongation of $X_{n+1}$ is: $x \p_{x} + (ry + x^{r})\p_{y} + (rx^{r-1} + (r-1)y') \p_{y'} + ,...,+ ((r(r-1)...(r-n+1))x^{r-n} + (r-n)y^{(n)})\p_{y^{(n)}}$.\\

The equation is:
\be
y^{(n)} = K e^{-y^{(n-1)} \over (n-1)!}\,K\ne 0.\label{1.3}
\ee

$(26, n+1), r = n-3$, $n\geq 4$: $X_{1} = \p_{x}$, $X_{2} = \p_{y}$, $X_{3} = x\p_{x}$, $X_{4} = y \p_{y}$, $X_{5} = x \p_{y}$,..., $X_{n+1} = x^{n-3}\p_{y}$.\\

Here  the equation turns out to be
\be
y^{(n)} = K \frac{(y^{(n-1)})^{2}}{y^{(n-2)}},\, K\ne0,n/(n-1).\label{1.4}
\ee

If $K=n/(n-1)$, then there is one more symmetry $X=x^2\p x+(n-3)xy\p y$ as discussed in Section 4.\\

$(27, n+1), r = n-3$, $n\geq 4$: $X_{1} = \p_{x}$, $X_{2} = \p_{y}$, $X_{3} = 2x\p_{x} + ry \p_{y}$, $X_{4} = x^{2} \p_{x} + rxy \p_{y}$, $X_{5} = x \p_{y}$,..., $X_{n+1} = x^{n-3}\p_{y}$.\\

The $n$th prolongation of $X_{3}$ and $X_{4}$ are: $X_{3} + \sum_{k=1}^{n}(r-2k)y^{(k)}\p_{y^{(k)}}$ and $X_{4} + \sum_{k=1}^{n}(k(r-k+1)y^{(k-1)} + x(r-2k)y^{(k)})\p_{y^{(k)}}$,
respectively.\\

The equation is
\be
y^{(n)} = \frac{n}{n-1}\frac{(y^{(n-1)})^{2}}{y^{(n-2)}} + K(y^{(n-2)})^{\frac{n+3}{n-1}},\,K\ne0.\label{1.5}
\ee
\\

If $K=0$, then there is one more symmetry $X=y\p y$ as pursued in Section 4.\\

$(28, n+1), r = n-4$, $n\geq5$: $X_{1} = \p_{x}$, $X_{2} = \p_{y}$, $X_{3} = x \p_{x}$, $X_{4} = y \p_{y}$, $X_{5} = x^{2}\p_{x} + (n-4)xy\p_{y}$, $X_{6} = x\p_{y}$, ..., $X_{n+1} = x^{n-4}\p_{y}$.\\

The $n$th order invariant equation is:
\be
y^{(n)} = (y^{(n-2)})^{3}(y^{(n-3)})^{-2}\bigg[\frac{n(3(n-2)K_{1}-2n+2)}{(n-2)^{2}}+K((n-2)K_{1}-(n-1))^{\frac{3}{2}}\bigg],
\ee
where $K_{1} = y^{(n-3)}y^{(n-1)}(y^{(n-2)})^{-2}$.

\subsection{Linearizable Equations for Higher Order}

Here we consider linearization for higher order $n\geq3$ equations. We demonstrate how one obtains the linear form. The reader is also referred to \cite{Mah2}. \\

$(21, n+1), r = n-1$: $X_{1} = \p_{y}$, $X_{2} = y\p_{y}$, $X_{3} = \x_{1}(x)\p_{y}$, ..., $X_{n+1} = \x_{r}(x)\p_{y}$.\\

By introducing coordinates: $\ov{x} = \x_{1}(x)$, $\ov{y} = y$ and ignoring the bars, the generators of this algebra can be transformed to $X_{1} = \p_{y}$, $X_{2} = y\p_{y}$, $X_{3} = x\p_{y}$, $X_{4} = \x_{2}(x)\p_{y}$,..., $X_{n+1} = \x_{r}(x)\p_{y}$.\\

Consider the general form of an $n$th order ODE as \eqref{0.1}.
For such an equation to admit $X_{1}$ and $X_{3}$,  $H$ must be independent of $y$ and $y'$. Invoking $X_{4}$ on this equation, the general form of $H$ can be determined as
\be
H = \frac{\x_{2}^{(n)}}{\x_{2}^{(n-1)}}y^{(n-1)} + H_{1}(x, \pi_{1}, \pi_{2} ,..., \pi_{n-3})\label{1.12}
\ee
where
\bc
$\pi_{k} = y^{(k+2)} - \dfrac{\x_{2}^{(k+2)}}{\x_{2}^{(k+1)}}y^{(k+1)},\hspace{1cm} k= 1, 2, ... ,n-3.$\\
\ec
Now imposing $X_{5}$ on this, the general form of $H_{1}$ is  found to be
\be
H_{1} = \frac{\pi^{0}}{\pi^{1}}\pi_{n-3} + H_{2}(x, \psi^{1}, \psi^{2}, ..., \psi^{n-4})\label{1.13}
\ee
where
\bc
$\pi^{k} = \x_{3}^{(n-k)} - \x_{3}^{(n-k-1)}\dfrac{\x_{2}^{(n-k)}}{\x_{2}^{(n-k-1)}}, \hspace{1cm} k= 0, 1, 2, ..., n-3.$\\
\ec
and
\bc
$\psi^{k} = \pi_{k+1} - \dfrac{\pi^{n-k-3}}{\pi^{n-k-2}}\pi_{k}, \hspace{1cm} k = 1, 2, 3, ..., n-4.$
\ec
\vspace{1cm}
Similarly, for an equation to be invariant under each $X_{i}$, for $i \geq 5$, the functions $H_{i-2}$s that are arbitrary functions of their $n-i-1$ arguments, can be determined. As a result, we can write the general form of an $n$th order ODE invariant under $X_{1}$,$X_{3}$,..., $X_{n+1}$ as
\be
y^{(n)} = \sum_{i=2}^{n-1}A_{i}(x)y^{(i)} + H_{n-2}(x).\label{1.14}
\ee\\

The invariance under $X_{2}$ now easily implies that $H_{n-2}$ must vanish giving us the following homogenous equation whose symmetry algebra is $(21, n+1)$:
\be
y^{(n)} = \sum_{i=2}^{n-1}A_{i}(x)y^{(i)}.\label{1.15}
\ee
One notices that since this equation is invariant under $X_{1}$, $X_{3}$,..., $X_{n+1}$, each $\x_{i}$ for $i = 2, 3, ... ,n-1$ must be an independent solution of equation \eqref{1.15}. In general this proves the following proposition.\\

\bp\label{pro1}
$(21, n+1)$ is a symmetry algebra of the $n$th order linear homogenous equation
\bc
$y^{(n)} = \sum_{i=2}^{n-1}A_{i}(x)y^{(i)}$,\\
\ec
such that each $\x_{i}$ for $ i = 1, 2, ..., n-1$ together with the number 1 form a set of independent solutions of this equation. Moreover, the functions $A_{i}$s are determined by solving the system of homogenous equations
\bc
$\x_{k}^{(n)} = \sum_{i=1}^{n-1}A_{i}(x)\x_{k}^{(i)}, \hspace{1cm} k = 1, 2, ... , n-1$.
\ec
\ep

\section{(n+2)-Dimensional Algebras}
\subsection{Nonlinear Equations}

For the $n+2$ dimensional algebras, $(20, n+2)$, $(21,n+2)$ and $(22,n+2)$ are clearly not admissible algebras. The linearization case $(23,n+2)$ is looked at in 4.2.\\

$(24, n+2), r = n-1$, $n\geq2$: $X_{1} = \p_{x}$, $X_{2} = \p_{y}$, $X_{3} = x \p_{x} + \alpha y \p_{y}$, $X_{4} = x \p_{y}$, $X_{5} = x^{2} \p_{y}$, ..., $X_{n+2} = x^{n-1} \p_{y}$.
\be
\Lambda_{L} = 1\cdot2!\cdot3!...(n-1)!(\alpha-n)y^{(n)} .\label{1.9}
\ee\\

$(25, n+2), r = n$, $n\geq1$: $X_{1} = \p_{x}$, $X_{2} = \p_{y}$, $X_{3} = x \p_{y}$, $X_{4} = x^{2}\p_{y}$ , ..., $X_{n+1} = x^{n-1} \p_{y}$, $X_{n+2} = x \p_{x} + (ry + x^{r})\p_{y}$.\\
\be
\Lambda_{L} = 1\cdot2!\cdot3!....(n-2)!\cdot(n-1)!\cdot n!.\label{1.6}
\ee\\

$(26, n+2), r = n-2$, $n\geq3$: $X_{1} = \p_{x}$, $X_{2} = \p_{y}$, $X_{3} = x\p_{x}$, $X_{4} = y \p_{y}$, $X_{5} = x \p_{y}$,..., $X_{n+2} = x^{n-2}\p_{y}$.
\be
\Lambda_{L} = 1\cdot2!\cdot3!...(n-3)!\cdot(n-2)!y^{(n-1)}y^{(n)}.\label{1.10}
\ee\\

$(27, n+2), r = n-2$, $n\geq3$: $X_{1} = \p_{x}$, $X_{2} = \p_{y}$, $X_{3} = 2x\p_{x} + ry \p_{y}$, $X_{4} = x^{2} \p_{x} + rxy \p_{y}$, $X_{5} = x \p_{y}$,..., $X_{n+2} = x^{n-2}\p_{y}$.
\be
\Lambda_{L} = 1\cdot2!\cdot3!...(n-2)!\cdot n^{2}(y^{(n-1)})^{2}.\label{1.8}
\ee\\

$(28, n+2), r = n-3$, $n\geq4$: $X_{1} = \p_{x}$, $X_{2} = x\p_{x}$, $X_{3} = y \p_{y}$, $X_{4} = x^{2}\p_{x} + rxy \p_{y}$, $X_{5} = \p_{y}$, $X_{6} = x \p_{y}$,..., $X_{n+2} = x^{n-3}\p_{y}$.\\

Equation:\be
y^{(n)} = \frac{n}{n-1} \frac{(y^{(n-1)})^{2}}{y^{(n-2)}}.\label{1.6}
\ee

Here we have only one case, the last,  which constitute an invariant equation with $n+2$ dimensional symmetry algebra. The rest do not form an equation or are singular equations with maximal algebra.\\

Next we need to again deal with linearization.

\subsection{Linearizable Higher Order Equations}

Here we obtain the form for the reduced linear equation that results from linearizability
(see also \cite{Mah2}).\\

$(23, n+2), r=n$: $X_{1} = \e_{1}(x) \p_{y}$, $X_{2} = \e_{2}(x) \p_{y}$, ..., $X_{n} = \e_{n}(x) \p_{y}$, $X_{n+1} = y \p_{y}$, $X_{n+2} = \p_{x}$.\\

From Proposition \ref{pro1}, it is easy to show that the general form of an $n$th order equation invariant under the first $n+1$ generators is the following linear homogenous equation:
\be
y^{(n)} = \sum_{i=0}^{n-1}A_{i}(x)y^{(i)}.\label{1.16}
\ee
For such an equation to admit $X_{n+2}$ the coefficients $A_{i}$s must be constant for each $i = 0, 1, 2, ..., n-1$. Note also that $\e_{i}$s for $i= 1, 2, ..., n$ form an independent set of solutions for this constant coefficient equation.\\

We can determine the $A_{i}$s as in Proposition \ref{pro1}  for known $\e_{i}(x)$s. However, note that these coefficients can also be determined by the roots of the corresponding characteristic polynomial equation.\\

We first consider the case when all the roots are real and distinct say, $\al_{i}$, \ $i = 1, 2, 3, ..., n$.\\

The fundamental solutions of equation \eqref{1.16} are:
\bc
$ \e_{i}(x) = e^{\al_{i}x}$, \hspace{1cm} for $i = 1, 2, 3, ..., n$.
\ec
The constants $A_{i}$s which depend on $\al_{i}$s can then be derived by solving the system
\be
VX = B, \label{1.17}
\ee
where $V$ is an $n\times n$ Vandermonde matrix in $(\al_{1}, \al_{2}, ... , \al_{n})$,
\bc
$X = [A_{0}, A_{1}, ... ,A_{n-1}]^{T}$ and $B = [\al_{1}^{n}, \al_{2}^{n}, ..., \al_{n}^{n}]^{T}$.
\ec
The system given in \eqref{1.17} can be solved by Crammer's rule, i.e. we have by Crammer's rule,
\bc
$A_{i}(\al_{1}, \al_{2}, ... , \al_{n}) = \dfrac{\det(f_{i})}{\det(V)}, \hspace{1cm} i = 0, 1, 2, ..., n-1$,
\ec
where $f_{i}$ is the matrix obtained by replacing the $(i+1)$th column in $V$ by $B$. First, we state the following lemma whose proof is elementary and thus omitted.
\blm
Let $V$ be a Vandermonde matrix in $(\al_{1}, \al_{2}, ..., \al_{n})$, then the determinant of $V$ is
\bc
 $\det(V) = \prod_{1\leq i < j \leq n}(\al_{i} - \al_{j})$.
 \ec
\elm
As an example, we find $A_{n-1}$. For this we need to find the determinant of $f_{n}$. We prove by induction on $n$ that
\bc
$A_{n-1}(\al_{1}, \al_{2},\cdots, \al_{n}) = \sum_{k=1}^{n}\al_{k}$.
\ec
Actually, we show that
\bc
$\det(f_{n}) = A_{n-1}\det(V)$.
\ec
For this we write the matrix $f_{n}$ as below:
\bc
$f_{n} =
\bpmat
1 & \al_{1} & \al_{1}^{2} & \cdots & \al_{1}^{n-2} & \al_{1}^{n} \\
1 & \al_{2} & \al_{2}^{2} & \cdots & \al_{2}^{n-2} & \al_{2}^{n} \\
\vdots  & \vdots  & \ddots & \vdots  \\
1 & \al_{n} & \al_{n}^{2} & \cdots & \al_{n}^{n-2} & \al_{n}^{n}
\epmat$
\ec
To find the determinant of this matrix we apply row operations to make all the entries in the first column zero except the first one.\\

$\Rightarrow$
\bc
$\det(f_{n})=
\bvmat
1 & \al_{1} & \al_{1}^{2} & \cdots & \al_{1}^{n-2} & \al_{1}^{n} \\
0 & \al_{1} - \al_{2} & \al_{1}^{2} - \al_{2}^{2} & \cdots & \al_{1}^{n-2} - \al_{2}^{n-2} & \al_{1}^{n} - \al_{2}^{n} \\
\vdots  & \vdots  & \ddots & \vdots  \\
0 & \al_{1} - \al_{n} & \al_{1}^{2} - \al_{n}^{2} & \cdots & \al_{1}^{n-2} - \al_{n}^{n-2}& \al_{1}^{n} - \al_{n}^{n}
\evmat$,
\ec
where by the vertical bars we denote the determinant of the matrix.\\
$\Rightarrow$
\bc
$\det(f_{n})=
\bvmat
\al_{1} - \al_{2} & \al_{1}^{2} - \al_{2}^{2} & \cdots & \al_{1}^{n-2} - \al_{2}^{n-2} & \al_{1}^{n} - \al_{2}^{n} \\
\al_{1} - \al_{3} & \al_{1}^{2} - \al_{3}^{2} & \cdots & \al_{1}^{n-2} - \al_{3}^{n-2} & \al_{1}^{n} - \al_{3}^{n} \\
\vdots  & \vdots  & \ddots & \vdots  \\
\al_{1} - \al_{n} & \al_{1}^{2} - \al_{n}^{2} & \cdots & \al_{1}^{n-2} - \al_{n}^{n-2}& \al_{1}^{n} - \al_{n}^{n}
\evmat$
\ec
$\Rightarrow$
\bc
$\det(f_{n}) = \prod_{k=2}^{n}(\al_{1} - \al_{k})
\bvmat
1 & \al_{1} + \al_{2} & \al_{1}() + \al_{2}^{2} & \cdots & \al_{1}() + \al_{2}^{n-3} & \al_{1}(\al_{1}() + \al_{2}^{n-2}) + \al_{2}^{n-1} \\
1 & \al_{1} + \al_{3} & \al_{1}() + \al_{3}^{2} & \cdots & \al_{1}() + \al_{3}^{n-3} & \al_{1}(\al_{1}() + \al_{2}^{n-2}) + \al_{3}^{n-1} \\
\vdots  & \vdots  & \ddots & \vdots  \\
1 & \al_{1} + \al_{n} & \al_{1}() + \al_{n}^{2} & \cdots & \al_{1}() + \al_{n}^{n-3} & \al_{1}(\al_{1}() + \al_{n}^{n-2}) + \al_{n}^{n-1}
\evmat$
\ec
where in the brackets we have the corresponding entries of the previous columns in each successive step.\\
$\Rightarrow$
\bc
$\det(f_{n}) = \prod_{k=2}^{n}(\al_{1} - \al_{k})
\bvmat
1 & \al_{2} & \al_{2}^{2} & \cdots & \al_{2}^{n-3} & \al_{1}(\al_{2}^{n-2}) + \al_{2}^{n-1} \\
1 & \al_{3} & \al_{3}^{2} & \cdots & \al_{3}^{n-3} & \al_{1}(\al_{2}^{n-2}) + \al_{3}^{n-1} \\
\vdots  & \vdots  & \ddots & \vdots  \\
1 & \al_{n} & \al_{n}^{2} & \cdots & \al_{n}^{n-3} & \al_{1}(\al_{n}^{n-2}) + \al_{n}^{n-1}
\evmat$
\ec
$\Rightarrow$
\bc
$\det(f_{n}) = \prod_{k=2}^{n}(\al_{1} - \al_{k})\Big(\al_{1}
\bvmat
1 & \al_{2} & \al_{2}^{2} & \cdots & \al_{2}^{n-3} & \al_{2}^{n-2} \\
1 & \al_{3} & \al_{3}^{2} & \cdots & \al_{3}^{n-3} & \al_{2}^{n-2} \\
\vdots  & \vdots  & \ddots & \vdots  \\
1 & \al_{n} & \al_{n}^{2} & \cdots & \al_{n}^{n-3} & \al_{n}^{n-2}
\evmat$ + $\bvmat
1 & \al_{2} & \al_{2}^{2} & \cdots & \al_{2}^{n-3} &  \al_{2}^{n-1} \\
1 & \al_{3} & \al_{3}^{2} & \cdots & \al_{3}^{n-3} &  \al_{3}^{n-1} \\
\vdots  & \vdots  & \ddots & \vdots  \\
1 & \al_{n} & \al_{n}^{2} & \cdots & \al_{n}^{n-3} &  \al_{n}^{n-1}
\evmat$\Big)
\ec
The first matrix is the Vandermonde matrix in $(\al_{2}, \al_{2},\cdots, \al_{n})$ and the second matrix is $f_{n}$ in $(\al_{2}, \al_{2},\cdots, \al_{n})$. By induction we now have
\bc
$\det(f_{n}) = \prod_{k=1}^{n}(\al_{1} - \al_{k})\Big(\al_{1}(\det(V(\al_{2}, \al_{3}, \cdots, \al_{n}))) + (\sum_{k=2}^{n}\al_{k})(\det(V(\al_{2}, \al_{3}, \cdots, \al_{n})))\Big)$
\ec
$\Rightarrow$
\bc
$\det(f_{n}) = \sum_{k=1}^{n}\al_{k}(\det(V(\al_{1}, \al_{2}, \cdots \al_{n})))$,
\ec
which proves the assertion.\\

In general we have the following lemma:
\blm \label{1.2}
Let
\bc
$A_{i} = \dfrac{\det(f_{i})}{\det(V)},$
\ec where $V$ and $f_{i}$ are defined as in the above discussion. Then $A_{i}$ can be determined as
\bc
$A_{i} = (-1)^{i}\sum_{1 \leq k_{1} < k_{2} < \cdots k_{i+1} \leq n}\al_{k_{1}}\al_{k_{2}} \cdots \al_{k_{i+1}}$.
\ec
\elm
The proof of this can be carried out by using induction on $n$ and $i$ simultaneously and therefore is omitted. The $A_{n-1}$ above was derived by using the technique of the proof of this lemma.\\

Now we consider the case when $2k$ number of the roots are complex and the rest real, i.e. $n -2k$ of them. If we show the complex roots as $\al_{j} = a_{j} + i b_{j}$ and $\ov{\al_{j}} = a_{j} - i b_{j}$ for $ j = 1, 2, \cdots k$ and the real roots as $\al_{2k+1}, \al_{2k+2} \cdots \al_{n}$. Then the fundamental set of solutions of equation \eqref{1.16} with constant coefficients can be taken as $\e_{j}(x) = e^{a_{j}(x)}\cos(b_{j}x)$ for $j = 1, 2, \cdots k$,  $\e_{j}(x) = e^{a_{j}x)}\sin (b_{j}x)$ for $ j =k+1 , k+2, \cdots 2k$ and the rest  as $e^{\al_{j}(x)}$s which correspond to the real roots. The constants $A_{i}$s can be derived in exactly the same way. Thus, we actually prove the following proposition.
\bp
The generators given in $(23, n+2)$ form a Lie algebras if and only if the $\e_{i}$s form a fundamental set of solutions for the constant coefficient equation given as
\be
y^{(n)} = \sum_{i=0}^{n-1}A_{i}y^{(i)}.\label{1.18}
\ee
Moreover, the constants can be determined as in lemma \ref{1.2}.
\ep

\section{Fundamental Invariants}

In this section we now classify all the fundamental invariants for five and higher dimensions. Note that the algebras $(20,n)$ and $(21,n)$ are not admissible.\\

$(5, 5)$:  The generators of this algebra are: $X_{1} = \p_{x}$, $X_{2} = \p_{y}$, $X_{3} = x \p_{x} - y \p_{y}$, $X_{4} = y \p_{x}$, $X_{5} = x \p_{y}$.\\

This is a five-dimensional algebra, hence $r = 5$. We find the
3rd prolongations of the generators of this algebra. Then the Lie determinant, which  is the determinant of the matrix
\bc
$M =
\bpmat
1 & 0 & 0 & 0 & 0 \\
0 & 1 & 0 & 0 & 0 \\
0 & x & 1 & 0 & 0 \\
x & -y & -2y' & -3y'' & -4y'''\\
y & 0 & -y'^{2} & -3y'y'' & -(3y''^{2} + 4y'y'''),
\epmat$
\ec
 is then
\bc
$\Lambda_{L}(M) = 9y''^{2}$.
\ec
This shows that $y'' = 0$ is the only equation of order $ \leq 3$ invariant under this algebra.\\

The fundamental differential invariants can then be found in the usual way. For the given algebra we already know a fourth order invariant which is
\bc
$\pi_{1} = y^{(4)}y''^{\frac{-5}{3}} - \frac{5}{3}y'''^{2}y''^{-8/3}$.
\ec
To find the 5th order differential invariant we can solve the system of linear partial differential equations
\be
 pr_{5}(X_{i})(y^{(5)} - H)|_{eq} = 0, \hspace{0.5cm} {\rm for} \  i = 1, 2, \cdots, 5, \label{1.20}
\ee
where $pr_{n}(X)$ is the  $n$th prolongation of $X$.\\

Solving this system we derive the 5th order equation invariant under this algebra to be
\be
y^{(5)} = \frac{-40}{9}y'''^{3}y''^{-2} + 5y'''y^{(4)}y''^{-1} + y''^{2}H(\pi_{1}),\label{1.21}
\ee
where $\pi_{1}$ is the fourth order invariant of this algebra and $H$ is an arbitrary function of its argument.
We can thus write the fifth order differential invariant to be
\bc
$\pi_{2} = (y''^{2}y^{(5)} + \frac{40}{9}y'''^{3} - 5y''y'''y^{(4)})/y''^{4}.$
\ec
Here $\pi_{1}$ and $\pi_{2}$ are the fundamental differential invariants for this algebra. All higher order differential invariants can be deduced by Lie's invariant differentiation:
$\pi_{n+1} = {D_{x}(\pi_{n})}/{D_{x}(\pi_{n-1})}$
or by using invariant derivative operator ${\cal D}=y''^{-1/3}D_x$ so that higher order equations are given by
\be
{\cal D}^{n-4}\phi_1=H(\phi_1,\ldots,{\cal D}^{n-5}\phi_1),\,n\ge5, \label{1.22}
\ee
where $D_{x}$ is the total derivative operator.\\

$(15, 5)$: $X_{1} = \p_{x}$, $X_{2} = \p_{y}$, $X_{3} = x \p_{x}$, $X_{4} = y \p_{y}$, $X_{5} = x^{2} \p_{x}$.\\

The fundamental invariants are
\bc
$\pi_{1} = \dfrac{y'^{2}y^{(4)} - 6(y'y''y''' - y''^{3})}{(2y'y''' - 3y''^{2})^{\frac{3}{2}}}$,\\
$\pi_{2} = \dfrac{3y'^{3}y^{(5)} - 10y'^{2}y''y^{(4)} -50y'^{2}y'''^{2}+ 240y'y''^{2}y''' - 180y''^{4}}{(2y'y''' - 3y''^{2})^{2}}$
\ec
with the 5th order equation as $\phi_2=H(\phi_1)$ and with ${\cal D}=y'(2y'y'''-3y''^2)^{-1/2}D_{x}$ we can invoke (\ref{1.22}) for higher order equations.\\

$(6, 6)$: $X_1 = \p_x$, $ X_2 = \p_y$, $X_3 = x \p_x$, $X_4 = y \p_x$, $X_{5} = x \p_y$, $X_{6} = y \p_y$.\\

The generators $X_1$, $X_2$, $X_5$ and $X_6$ result in the invariants
$$
K_1=3y''^{-1}y^{(4)} - 5y'''^{2}y''^{-2},
$$
$$
K_2= 9y''^{-1}y^{(5)} - 45y''^{-2}y'''y^{(4)} + 40y''^{-3}y'''^{3}.
$$
Writing $X_4$ in terms of $K_1$ and $K_2$ we find
$$
X_4=2K_1\p_{K_1}+3K_2\p_{K_2}
$$
which provides  the 5th order invariant of $(6,6)$
\bc
$\phi_1=K_1^{-3/2}K_2$,\\
$= \dfrac{9y''^{2}y^{(5)} - 45y''y'''y^{(4)} + 40y'''^{3}}{(3y''y^{(4)} - 5y'''^{2})^{\frac{3}{2}}}$.
\ec
The invariant differentiation operator is ${\cal D}=K_1^{-1/2}D_x$ and hence one can derive a sixth order invariant
\bc
$\pi_2 = \dfrac{(9y''^{3}y^{(6)} - 63y''^{2}y'''y^{(5)} + 105y''y'''^{2}y^{(4)} -35y'''^{4})}{(3y''y^{(4)} - 5y'''^{2})^{2}}$.
\ec

The 5th order invariant equation is $\phi_1=C$, $C$ a constant and with ${\cal D}=y''(3y''y^{(4)}-5y'''^2)^{-1/2}D_x$, we have
\be
{\cal D}^{n-5}\phi_1=H(\phi_1,\ldots,{\cal D}^{n-6}\phi_1),\,n\ge6,\label{d11}
\ee
for higher order equations. Here the 6th order equation is $\phi_2=H(\phi_1)$ with $\phi_2$ as above and higher order equations are deduced by invariant differentiation as stated.\\

Note that $K_1=0$ is the fourth order singular invariant equation with this algebra.\\

$(16,6)$: $X_1 = \p_x$, $ X_2 = \p_y$, $X_3 = x \p_x$, $X_4 = y \p_y$, $X_{5} = x^2 \p_x$, $X_{6} = y^2 \p_y$.\\

Here the invariants, using $X_1$, $X_2$, $X_4$ and $X_6$, are
$$
K_1=y'''y^{-1} - \frac32y'^{-2}y''^{2},
$$
$$
K_2= y'^{-1}y^{(4)} +3y'^{-3}y''^3 -4y'^{-2}y''y'''.
$$
$$
K_3=y'^{-1}y^{(5)}+5y'^{-3}y''^2y'''-5y'^{-2}y''y^{(4)}.
$$
Utilising $X_5$ in $(K_1,K_2,K_3)$ space, we end up with
$$
X_5=2xK_1\p_{K_1}+(3xK_2+2K_1)\p_{K_2}+(4xK_3+5K_2)\p_{K_3}.
$$
Hence, a 5th order invariant transpires as
$$
\phi_1=K_1^{-3}(5K_2^2-4K_1K_3).
$$
The invariant differentiation operator is ${\cal D}=K_1^{-1/2}D_x$ and higher order invariant equations are as in (\ref{d11}).\\

Here $K_1=0$ is the third order singular invariant equation admitting this algebra.\\

$(7,6)$: $X_1 = \p_x$, $ X_2 = \p_y$, $X_3 = x \p_x+y\p_y$, $X_4 = y \p_x-x\p_y$, $X_{5} = (x^2 -y^2)\p_x+2xy\p_y$, $X_{6} = 2xy\p_x+(y^2-x^2) \p_y$.\\

The operators $X_1$ to $X_4$ gives rise to the invariants
$$
K_1=(1+y'^2)y''^{-2}y''' - 3y',
$$
$$
K_2= (1+y'^2)^2y''^{-3}y^{(4)} +15y'^{2} -10y'y''^{-2}y'''(1+y'^2),
$$
$$
K_3=(1+y'^2)^3y''^{-4}y^{(5)}-15(1+y'^2)^2y'y''^{-3}y^{(4)}-105y'^{3}
$$
$$
+105(1+y'^2)y'^2y''^{-2}y'''-10(1+y'^2)^2y'y''^{-4}y'''^2.
$$

Now $X_5$ in $(K_1,K_2,K_3)$ space yields the two equations
$$-2K_1\p_{K_2}F+(15-5K_2)\p_{K_3}F=0,$$
$$-2K_1\p_{K_1}F+(9-3K_2)\p_{K_2}F+(40K_1-4K_3)\p_{K_3}F=0.$$

The invariant deduced is of order five (which is admitted by $X_6$ as well since
$[X_4,X_5]=X_6$):

$$
\phi_1=30K_1^3K_2-5K_1^3K_2^2+4K_1^4K_3+45K_1^3+16K_1^5.
$$
The operator of invariant differentiation is ${\cal D}=(1+y'^2)y''^{-1}K_1^{-1/2}D_{x}$
and higher order equations are given by (\ref{d11}). \\

Here $K_1=0$ is the third order singular invariant equation.\\

$(8,8)$: $X_1 = \p_x$, $ X_2 = \p_y$, $X_3 = x \p_y$, $X_4 = y \p_y$, $X_{5} = y\p_x$, $X_{6} = x\p_x$, $X_7=x^2\p_x+xy \p_y$, $X_8=xy\p_x+y^2\p_y$.\\

The seventh order invariant is
$$\phi=K_2^{-5/3}D_x^2 K_2-\frac76 K_2^{-8/3}(D_x K_2)^2-
y''^{-1}y'''K_2^{-5/3}D_x K_2$$
$$-y''^{-2}y'''^2K_2^{2/3}-\frac32y''^{-1}y^{(4)}K_2^{-2/3},
$$

where $K_2= 9y''^{-1}y^{(5)} - 45y''^{-2}y'''y^{(4)} + 40y''^{-3}y'''^{3}$  is the invariant of algebra $(6,6)$. The operator of invariant differentiation is ${\cal D}=K_2^{-1/3}D_x$ and higher order equations are obtained as
\be
{\cal D}^{n-7}\phi=H(\phi,\ldots,{\cal D}^{n-8}\phi),\,n\ge8.\label{d12}
\ee

For this algebra the singular invariant equation is $K_2=0$ which is fifth order.\\

$(22, n)$:\, $r = n-1$, $n\ge2$: $X_{1} = \p_{x}$, $X_{2} = \eta_{1}(x)\p_{y}$, \ldots, $X_{n} = \eta_{n-1}(x)\p_{y}$.\\

The $n$th order equation is
$$
 y^{(n)}+a_1y^{(n-1)}+\cdots+a_{n-1}y'=H(y^{(n-1)}+a_1y^{(n-2)}+\cdots+a_{n-1}y),
 $$
 where
 the $a_i$s are constants and the $\eta_i$s satisfy the linear $(n-1)$th order equation
 $$
 E_i\equiv \eta_i^{(n-1)}+a_1\eta_i^{(n-2)}+\cdots+a_{n-1}\eta_i=0,
 i=1,\ldots,n-1
 $$
 and its derived equation $D_x E_i=0$.\\



$(23, n):\, r = n-2$, $n\ge3$: $X_{1} = \p_{x}$, $X_{2} = y \p_{y}$, $X_{3} = \eta_{1}(x)\p_{y}$,...,$X_{n} = \eta_{n-2}(x)\p_{y}$.\\

The $n$th order equation that admits this algebra is
$$
 D_x^2\ln|y^{(n-2)}+a_1y^{(n-3)}+\cdots+a_{n-2}y|=H(D_x\ln|y^{(n-2)}+a_1y^{(n-3)}+\cdots+a_{n-2}y|),
 $$
 where
 the $a_i$s are constants and the $\eta_i$s satisfy the linear $(n-1)$th order equation
 $$
 \eta_i^{(n-2)}+a_1\eta_i^{(n-3)}+\cdots+a_{n-2}\eta_i=0,
 i=1,\ldots,n-2
 $$

$(24, n): r = n-3$, $n\ge4$: $X_{1} = \p_{x}$, $X_{2} = \p_{y}$, $X_{3} = x \p_{x} + \alpha y\p_{y}$, $X_{4} = x\p_{y}$,..., $X_{n} = x^{n-3}\p_{y}$.\\

Fundamental Invariants, if  $\alpha\ne n-1$, are
\begin{center}
  $\phi_{1} = {(y^{(n-1)})}^{\alpha-n+2}(y^{(n-2)})^{-(\alpha-n+1)}$, \hspace{0.6cm} $\phi_{2} = y^{(n)}(y^{(n-1)})^{-(\frac{\alpha - n}{\alpha-n+1})},$
\end{center}
and for $\alpha=n-1$
\begin{center}
  $\phi_{1}=y^{(n-1)}$,\hspace{0.6cm}
  $\phi_2=y^{(n)}y^{(n-2)}.$
 \end{center}

Thus the invariant equation for both cases is $\phi_2=H(\phi_1)$.\\

If $\alpha=n-1$, the invariant equation is
$$
y^{(n)}=(y^{(n-2)})^{-1}H(y^{(n-1)}).
$$

$(25, n): r = n-2$, $n\ge3$: $X_{1} = \p_{x}$, $X_{2} = \p_{y}$, $X_{3} = x \p_{y}$, $X_{4} = x^{2} \p_{y}, \cdots, X_{n-1} = x^{n-3} \p_{y}$, $ X_{n} = x \p_x+((n-2)y+x^{(n-2)}\p_{y}$. \\

The fundamental differential invariants are
\bc
$\pi_{1} = y^{(n-1)}e^{\frac{y^{(n-2)}}{(n-2)!}}, \ \ \pi_{2} = y^{(n)}e^{\frac{2y^{(n-2)}}{(n-2)!}}$
\ec
and invariant equation as before $\phi_2=H(\phi_1)$.\\

$(26, n)$, $r=n-4$, $n\ge5$: $X_{1} = \p_{x}$, $X_{2} = \p_{y}$, $X_{3} = x \p_{x}$, $X_{4} = y \p_{y}$, $X_{5} = x \p_{y}, \cdots, X_{n} = x^{n-4} \p_{y}$.\\

The fundamental differential invariants:
\bc
$\pi_{1} = \frac{y^{(n-3)}y^{(n-1)}}{(y^{(n-2)})^{2}}, \ \ \pi_{2} = \frac{(y^{(n-3)})^{2}y^{(n)}}{(y^{(n-2)})^{3}}$.
\ec

$(27, n)$, $r=n-4$, $n\ge5$: $X_{1} = \p_{x}$, $X_{2} = \p_{y}$, $X_{3} = 2x \p_{x} + y \p_{y}$, $X_{4} = x^{2} \p_{x} + xy \p_{y}$, $X_{5} = x \p_{y}, \cdots, X_{n} = x^{n-4} \p_{y}$. \\

The fundamental invariants:
\bc
$\pi_{1} = y^{(n-1)}(y^{(n-3)})^{\frac{n+2}{2-n}} - \dfrac{n-1}{n-2}(y^{(n-2)})^{2}(y^{(n-3)})^{\frac{-2n}{n-2}}$\\
$\pi_{2} = y^{(n)}(y^{(n-3)})^{\frac{n+4}{2-n}} + \dfrac{2n(n-1)}{(n-2)^{2}}(y^{(n-2)})^{3}(y^{(n-3)})^{\frac{3n}{2-n}} +\dfrac{3n}{2-n}y^{(n-1)}y^{(n-2)}(y^{(n-3)})^{\frac{2(n+1)}{2-n}}$.
\ec

$(28, n): r = n-5$, $n\ge6$: $X_1 = \p_x$, $ X_2 = \p_y$, $X_3 = x \p_y, \cdots, X_{n-3} = x^{n-5}\p_{y}$,  $X_{n-2} = x \p_x$, $ X_{n-1} = y \p_{y}$, $X_n = x^2 \p_x + rxy \p_y$.\\

The fundamental invariants:
\bc
$\pi_1 = \dfrac{(n-3)^{2}K_2 - 3(n-1)(n-3)K_1 +2(n-1)(n-2)}{
((n-3)K_1-(n-2))^{\frac{3}{2}}}$,\\
$\pi_2 = K_3{((n-3)K_1 -(n-2))}^{-2} +{((n-3)K_1 -(n-2))}^{-3}[-\frac32(n-3)K_2^2$\\
$+\frac12(5n-13)K_1K_2-3(n-1)K_1^3+\frac32(n-1)K_1^2-
2(n-2)K_2]$,
\ec
where
\bc
$K_{i} = \dfrac{y^{(n-3+i)}(y^{(n-4)})^{i}}{(y^{(n-3)})^{i+1}},\,i=1,2,3$.
\ec

\section{(n + 3)-Dimensional Algebras}

Scalar first order equations admit infinite number of point symmetries. A second order equation does not admit a five dimensional symmetry algebra, as is well-known. The only third order equations that admit  6 dimensional algebras are
\bc
$y''' = \dfrac{3y''^{2}}{2y'}$ \  and \  $y''' = \dfrac{3y'y''^{2}}{1+y'^{2}}$,
\ec
where the algebras are $\maf{sl}(2)\oplus \maf{sl}(2)$ and $\maf{so}(3, 1)$, respectively.\\

In \cite{SFHT}, it was shown that a fourth order equation either does not admit a 7 dimensional algebra or if it admits such an algebra, then it is not maximal. \\

For fifth order equations, the only equation which admits an
8 dimensional algebra is
\be
y^{(5)} = \frac{5y'''y^{(4)}}{y''} - \frac{40y'''^{3}}{9y''^{2}},\label{5th}
\ee
whose symmetry algebra is $\maf{sl}(3)$ as stated in the previous section.\\

We check all the possible algebras of dimension $n + 3$ for $n \geq 5$. Since an $n$th order linear  equation cannot have more than $n$ independent solutions, the algebras $(m, n+3)$ for $m = 20, \cdots, 25$ are not admitted. We discuss the remaining possible algebras:\\

$(26, n+3)$: $X_{1} = \p_{x}$, $X_{2} = \p_{y}$, $X_{3} = x \p_{x}$, $X_{4} = y \p_{y}$, $X_{5} = x \p_{y}, \cdots, X_{n+3} = x^{n-1} \p_{y}$.\\

The generators $X_{1}$, $X_{2}$ and $X_{5},  \cdots, X_{n+1}$ implies that the equation must be of the form
 \bc
$y^{(n)} = K$.
\ec
The $X_{4}$ then implies that $K$ must vanish. Then $X_{3}$ is  automatically admitted by this equation. However, the maximal symmetry algebra of this equation is $n + 4$. In the same way it can be shown easily that the algebras $(27, n+3)$ and $(28, n+3)$ are admitted by an $n$th order equation if and only if the equation is equivalent to $y^{(n)} = 0$ and hence these algebras are admitted by the equation but are not maximal. This discussion proves the following proposition.\\

\bp
For $n \geq 4$ an $n$th order ODE either does not admit an
$(n + 3)$-dimensional algebra or it does admit it but is not maximal with the exception that for $n = 5$, $\maf{sl}(3)$ is admitted by the equation given as \eqref{5th}.
\ep

In the next section we review $n=4$ and discuss $n = 5$ at some length. Then we mention how one can obtain a classification for $n > 5$.

\section{The Cases  $n\geq4$}
In this section we review the main aspects on the classification of 4th order ODEs and then discuss the cases $n\geq 5$. We consider the 5th order case in detail as an example to show how one can obtain the classification of ODEs of any higher order. \\

In general we can classify scalar ODEs into 3 subclasses as follows:\\

Subclass (1): $n$th order equations admitting $n+1$, $n+2$, $n+3$ and the maximal $n+4$ dimensional algebras.\\

Subclass (2): $n$th order equations admitting $n$ dimensional algebras. For this we have found the fundamental invariants of $n$ dimensional algebras which are of order $n-1$ and $n$.\\

Subclass (3): $n$th order equations admitting algebras of dimension lower than $n$.\\

Subclasses (1) and (2) are completed in this work for $n \geq 5$ and the algebra of dimension $n+4$ is already a well-known algebra since Lie with  corresponding equation $y^{(n)} =0$ that possesses this maximal dimension Lie algebra. \\

Subclass (3) can be performed in the following manner:\\

For algebras of dimension 1 and 2, this is easy and is given in Table  \ref{Table1.1} below. \\

{\bf The $n$th Order Equations Admitting
1 and 2 Dimensional Algebras for $n \geq 3$}\\

\begin{table}[h]
  \centering
\begin{tabular}{|c|c|c|}
  \hline
  Algebra & Generators & Invariant Equation \\
  \hline
  (9, 1) & $\partial_{x}$ & $y^{(n)} = H(y, y', y'', ..., y^{(n-1)})$ \\
  (10, 2) & $\partial_{x}$, $x \partial_{x}$ & $y^{(n)} = y'^{n}H(y,y''y'^{-2},..., y^{(n-1)}y'^{1-n})$ \\
  (20, 2) & $\partial_{y}$, $x \partial_{y}$ & $y^{(n)} = H(x, y'', ...y^{(n-1)})$ \\
  (22, 2) & $\partial_{x}$, $\partial_{y}$ & $y^{(n)} = H(y', y'', ..., y^{(n-1)})$ \\
  $A_{2,1}$ & $\partial_{x}$, $x\partial_{x}+y\partial_{y}$ & $y^{(n)} = y^{1-n}H(y', y''y, y'''y^{2},..., y^{(n-1)}y^{n-2})$ \\
  \hline
\end{tabular}
\caption{The $n$th Order Equations Admitting
1 and 2 Dimensional Algebras for $n \geq 3$}\label{Table1.1}
\end{table}

For any 3 dimensional algebra we already know the fundamental invariants of orders 2, 3 and 4 from the works of Gat \cite{Gat}, Mahomed and Leach \cite{FL3} as well as  \cite{SFHT}. Then by invariant differentiation we can find higher order invariants up to the required order, i.e. $n$. For this Lie's recursive formula
\begin{equation}
w_{k} = \frac{D_{x}(w_{k-1})}{D_{x}(w_{k-2})}, k = 5, 6, ..., n
\end{equation}
can be utilised, where $w_{k}$ is a $k$th order invariant. The general form of an $n$th order equation admitting this 3 dimensional algebra can be given as:
\begin{equation}\label{invariants}
F(w_{2}, w_{3}, ..., w_{n}) = 0
\end{equation}
or as
\begin{equation}{\label{7.3}}
w_{n} = H(w_{2}, w_{3}, ..., w_{n-1}).
\end{equation}

Another way is to use operator of invariant differentiation ${\cal D}=\lambda D_x$.
For example, if we know a second order invariant $\phi_1$ of a low dimensional algebra, then higher order equations for this algebra are of the form
\begin{equation}
{\cal D}^{n-2}\phi_1=H(\phi_1,\ldots,{\cal D}^{n-3} \phi_1),\quad n\geq3.
\end{equation}

In a similar way, this can be done for algebras of higher dimensions.\\

We first briefly review fourth order equations that possess 4 dimensional algebras in a compact way than in \cite{SFHT}. This is presented in the following Table \ref{Table1.2}.\\

\begin{table}[h]
  \centering
\begin{tabular}{|c|c|c|c|}
  \hline
  Algebra  & $\lambda$  & 3rd order Invariant $\phi_1$ & 4th Order ODE\\
  \hline
  (4, 4) & $(1+y'^2)y''^{-1}$ & $y''^{-2}y'''(1+y'^2)-3y'$& $y^{(4)}=y''^3(1+y'^2)^{-2}(15y'^2$\\
  &&&$+10\phi_1y'+H(\phi_1)$ \\
  (13, 4) & $y'y''^{-1}$ & $y'y''^{-2}y'''$ & $y^{(4)}=y'^{-2}y''^{3}H(\phi_1)$ \\
  (14, 4) & $y'^{-1}$ & $y'''y'^{-3}-\frac32y''^2y'^{-4}$ & $y^{(4)}=y'^4H(\phi_1)+6y'^2y''\phi_1$\\
  &&&$+3y'^{-2}y''^3$\\
  (19, 4) & $y^{1/2}y''^{-1/2}$ & $y^{1/2}y''^{-3/2}y'''+3y'y^{-1/2}y''^{-1/2}$ & $y^{(4)}=\frac43y''^{-1}y'''^2+y^{-1}y''^2H(\phi_1)$ \\
  (22,4) &1&$y'''+a_1y''+a_2y'+a_3y$&$D_x\phi_1=H(\phi_1)$\\
 & &$\eta_i$ solves $\phi_1=0$ for $a_i$s const. &\\
 (23,4) & 1& $D_x\ln|E|\equiv D_x\ln|y''+a_1y'+a_2y|$&$D_x\phi_1=H(\phi_1)$\\
 &&$\eta_i$ satisfy $E=0$ for $a_i$ const.&\\
 (24,4) & $y''^{1/( \alpha-2)}$, &${y'''}^{\alpha-2} y''^{3-\alpha}$&$y^{(4)}=y'''{^{\alpha-4\over \alpha-3}}H(\phi_1)$, $\alpha\neq3$\\
 &$y''$&$y'''$&$y^{(4)}=y''^{-1}H(y''')$\\
 &$y'''^{-1}$&$y''$&$y^{(4)}=y'''^2H(y'')$\\
 (25,4)&$\exp (y''/2)$&$y'''\exp (y''/2)$&$y^{(4)}=\exp(-y'')H(\phi_1)$\\
  \hline
\end{tabular}
\caption{The $4$th Order Equations Admitting
4 Dimensional Algebras}\label{Table1.2}
\end{table}
\newpage
There are multiple cases of (22,4) in \cite{SFHT}.  These are considered as one case in the Table  \ref{Table1.2}. Note that invariant differentiation of the invariants in the above Table \ref{Table1.2} will give fifth and higher order invariants. \\

We now consider fifth order invariant equations admitting algebras of dimension 5 and greater. These are  given in Table \ref{Table1.3} and \ref{Table1.4} and follow from our deliberations in the previous sections.\\

\subsection*{The $5$th order invariant ODEs that correspond to their symmetry algebras}

\begin{table}[h]
  \centering
\begin{tabular}{|c|c|}
   \hline
   5 D Algebra & 5th order canonical equation\\
  \hline
   (5, 5) & $y^{(5)} = \frac{-40}{9}y'''^{3}y''^{-2} + 5y'''y^{(4)}y''^{-1} + y''^{2}H(\pi_{1})$ \\
   &$\pi_{1} = y^{(4)}y''^{\frac{-5}{3}} - \frac{5}{3}y'''^{2}y''^{-8/3}$\\
   \hline
   (15, 5) &$ \dfrac{3y'^{3}y^{(5)} - 10y'^{2}y''y^{(4)} -50y'^{2}y'''^{2}+ 240y'y''^{2}y''' - 180y''^{4}}{(2y'y''' - 3y''^{2})^{2}}=H(\phi_1)$ \\
  &$\pi_{1} = \dfrac{y'^{2}y^{(4)} - 6(y'y''y''' - y''^{3})}{(2y'y''' - 3y''^{2})^{\frac{3}{2}}}$\\
  \hline
   (22,5) & $D_x(E_4(y))=H(E_4(y))$\\
   & $\eta_i$ is solution of $E_4(\eta_i)=0$, $i=1,\ldots,4$ for $a_i$ consts.\\
   \hline
  \end{tabular}
\end{table}
\newpage

\begin{table}
  \centering
  \begin{tabular}{|c|c|}
   \hline
   5 D Algebra & 5th order canonical equation\\
  \hline
  (23,5) & $D_x^2\ln|(E_3(y))|=H(D_x\ln|E_3(y))|$\\
   & $\eta_i$ is solution of $E_3(\eta_i)=0$, $i=1,\ldots,3$ for $a_i$ consts.\\
  \hline
   (24, 5) & $ y^{(5)}=(y^{(4)})^{\frac{\alpha - 5}{\alpha - 4}}H((y^{(4)})^{\alpha-3}(y^{(3)})^{4-\alpha})$, $\alpha\neq4$\\
   & $y^{(5)}=(y''')^{-1}H(y^{(4)})$, $\alpha=4$\\
  \hline
   (25, 5) & $ y^{(5)}=\exp({-\frac{2y'''}{3!}})H(y^{(4)}\exp(y'''/3!))$\\
   \hline
   (26, 5) & $y^{(5)}y''^{2}y'''^{-3}=H(y''y'''^{-2}y^{(4)})$\\
   \hline
   (27, 5) & $y^{(5)}y''^{-3} + \frac{40}{9}y'''^{3}y''^{-5}-5y^{(4)}y'''y''^{-4}=H(\phi_1)$\\
   &$\phi_1=y^{(4)}y''^{-7/3}-\frac43y'''^2y''^{-10/3}$, $H\neq K\phi_1^{3/2}$, $K$ const.\\
   \hline
   \end{tabular}\\
   \caption{The $5$th order invariant ODES that correspond to their 5 dimensional symmetry  algebras. Note that $E_n(z)\equiv z^{(n)}+a_1z^{(n-1)}+\cdots+a_{n}z$.}\label{Table1.3}
\end{table}

\begin{table}[h]
  \centering
  \begin{tabular}{|c|c|}
  \hline
  Higher Algebra & 5th order invariant equation\\
  \hline
   (6, 6) & $\dfrac{9y''^{2}y^{(5)} - 45y''y'''y^{(4)} + 40y'''^{3}}{(3y''y^{(4)} - 5y'''^{2})^{\frac{3}{2}}}=K$, $K$ const.\\
   (16,6) & $K_1^{-3}(5K_2^2-4K_1K_3)=K$, $K$ const.\\
   (7,6) & $ K_1^3(30K_2-5K_2^2+4K_1K_3+45+16K_1^2)=K$, $K$ const.\\
   (21,6) & $ y^{(5)}=\sum_{i=2}^{4}A_i(x)y^{(i)}$, $\xi_k$ satisfy
   $ \xi_k^{(5)}=\sum_{i=2}^{4}A_i(x)\xi_k^{(i)}$, $k=1,\ldots,4$\\
   (24,6) &$y^{(5)}=K(y^{(4)})^{\frac{\alpha - 5}{\alpha - 4}}$, $\alpha\neq4$\\
   (25, 6) & $ y^{(5)}=K\exp(-y^{(4)}/4!)$\\
   (26, 6) & $y^{(5)}=Ky'''^{-1}(y^{(4)})^{2}$\\
   (27, 6) & $y^{(5)}=K(y^{(3)})^2+\frac54(y^{(4)})^{2}y'''^{-1}$\\
   (28, 6) & $(\frac{y^{(5)}y''^{2}}{y'''^{3}} - \frac{5(9K_{1}-8)}{9})/(3K_{1}-4)^{3/2}=K$, $K$ const. \\
   &$K_{1} = y^{(4)}y''y'''^{-2}$\\
   (28, 7) & $y^{(5)}=\frac54(y^{(4)})^{2}y'''^{-1}$\\
   (8,8) & $9y''^{2}y^{(5)} - 45y''y'''y^{(4)} + 40y'''^{3}=0$\\
   (28, 9) & $y^{(5)} = 0$\\
   \hline
  \end{tabular}\\
  \caption{The $5$th order invariant ODEs that correspond to their higher symmetry algebras}\label{Table1.4}
\end{table}
%
%
%
%
%

One can proceed to classify 6th order ODEs as follows. The algebras are (6,6) given by
$\phi_2=H(\phi_1$, where $\phi_1=K_1^{-3/2}K_2$ and $\phi_2=K_1^{-1/2}D_x\phi_1$;
(16,6) with equation
$$K_1^{-1/2}D_x[K_1^{-3}(5K_2^2-4K_3K_1)]=H(K_1^{-3}(5K_2^2-4K_3K_1));$$ (7,6) with representative ODE $${\cal D}[ K_1^3(30K_2-5K_2^2+4K_1K_3+45+16K_1^2)]=H( K_1^3(30K_2-5K_2^2+4K_1K_3+45+16K_1^2)).$$
Also one can easily obtain the forms for the equations for the algebras (22,6) to (28,6). Altogether, there are hence 10 canonical forms for 6th order ODEs admitting 6 dimensional algebras. Moreover, for higher symmetries, there are 9 types given by (21,7), (24,7) to (28,7) as well as (23,8), (28,8) and
(28,9).\\

Likewise, if we investigate 7th order ODEs, the number of canonical forms are 7 given by
(22,7) to (28,7). Moreover, for higher symmetries, we end up with 10 types, viz. (21,8), (24,8) to (28,8), (23,9), (28,9), (28,11) and (8,8) given by the ODE
$$K_2^{-5/3}D_x^2 K_2-\frac76 K_2^{-8/3}(D_x K_2)^2-
y''^{-1}y'''K_2^{-5/3}D_x K_2$$
$$-y''^{-2}y'''^2K_2^{2/3}-\frac32y''^{-1}y^{(4)}K_2^{-2/3}=K.
$$\\

Now proceeding to 8th order ODEs, we have the 8 canonical forms
(22,8) to (28,8) as well as (8,8) given by
$${\cal D}w=K_2^{-13}D_x(K_2^{-5/3}D_x^2 K_2-\frac76 K_2^{-8/3}(D_x K_2)^2-
y''^{-1}y'''K_2^{-5/3}D_x K_2$$
$$-y''^{-2}y'''^2K_2^{2/3}-\frac32y''^{-1}y^{(4)}K_2^{-2/3})=H(w)
$$
where ${\cal D}=K_2^{-13}D_x$.\\

There are also 9 types of higher symmetries, viz. (21,9), (24,9) to (28,9), (23,10), (28,10)
and (28,12).\\

We have the generalization to order 9 and greater as follows:\\

For $n\geq9$, there are 7 canonical forms for $n$th order ODEs admitting $n$ symmetries. These are $(22,n)$ to $(28,n)$.\\

Also, for higher symmetries for $n$th order equations, $n\geq9$, there are 9 types, viz.
$(21,n+1)$, $(24,n+1)$ to $(28,n+1)$, $(23,n+2)$, $(28,n+2)$ and $(28,n+4)$ the maximal algebra.\\


We also remark here that the invariants above are verified by Maple code confirming that it is correct. Below is an example to show this.\\

\begin{exm}

$(28, 6)$: $X_{1} = \p_{x}$, $X_{2} = \p_{y}$, $X_{3} = x\p_{x}$, $X_{4} = x \p_{y}$, $X_{5} = x^{2}\p_{x}+xy\p_{y}$, $X_{6} = y \p_{y}$.\\

The fifth order differential invariant is:
\bc
$ \pi = (\dfrac{y^{(5)}y''^{2}}{y'''^{3}} - \dfrac{5(9K_{1}-8)}{9})/(3K_{1}-4)^{\frac{3}{2}}$,  \\ $K_{1} = \dfrac{y^{(4)}y''}{y'''^{2}}$.
\ec

Using the Maple code we get the differential invariant :
\bc
$\pi = (1/9)(9y''^2y^{(5)}-45y''y'''y^{(4)}+40y'''^3)/(3y''y^{(4)}-4y'''^2)^{\frac{3}{2}}$
\ec
which is equivalent to the above invariant computed by hand.
\end{exm}

\section*{Appendix}
\subsection*{A. Some Fourth Order Invariants in \cite{SFHT}}
Here we provide the corrected invariants that had inadvertent  errors in  \cite{SFHT}.\\

(1, 3): $X_{1} = \p_{x}$, $X_{2} = \p_{y}$, $X_{3} = (\alpha x + y)\p_{x} + (\alpha y-x)\p_{y}$\\

The second and third order invariants are:
\begin{center}
$w_{2} = y''(1+y'^{2})^{\frac{-3}{2}}e^{-\alpha \arctan(y')}$, $w_{3} = \dfrac{y'''(1+y'^{2})}{y''^{2}} - 3y'$.
\end{center}
The fourth order invariant is
\begin{center}
$w = \lambda D_{x}(w_{3})$ where $\lambda = (1+y'^{2})^{\frac{-1}{2}}e^{-\alpha \arctan(y')}$
\end{center}
and simplifies as $w/w_2$ which is
\begin{center}
$
w_4=(y'^2+1)^2y''^{-3}y^{(4)}-2y''^{-^4}y'''{^2}(y'^2+1)^2+2y'y''^{-2}y'''(y'^2+1)-3(y'^2+1)$
\end{center}

(2, 3):
$X_{1} = \p_{y}$,  $X_{2} = x\p_{x} + y\p_{y}$, $X_{3} = 2xy\p_{x} + (y^{2} - x^{2})\p_{y}$\\

We know the second and third order invariants from the works \cite{FL3} and \cite{Gat} which are
\begin{center}
$w_{2} = xy''(1+y'^{2})^{\frac{-3}{2}} - y'(1+y'^{2})^{\frac{-1}{2}}$, $w_{3} = \dfrac{x^{2}y'''}{(1+y'^{2})^{2}} - \dfrac{-3x^{2}y'y''^{2}}{(1+y'^{2})^{3}}$
\end{center}
The fourth order invariant can then be found by invariant differentiation which is:
\begin{center}
$w_{4} = \lambda D_{x}(w_{3})$.
\end{center}
This can also be found once we know a suitable $\lambda$ which in this case is $\lambda = x(1+y'^{2})^{\frac{-1}{2}}$. The explicit form of this fourth order invariant is given below which is found using Maple. Moreover, the general form of a fourth order equation invariant under this algebra can also be written using the equation \eqref{7.3}.\\

The fourth order invariant:
\begin{center}
$\dfrac{1}{(y'^2+1)^(\frac{9}{2})}((15y'^2y''^3-10y'y'''(y'^2+1)y''+y^{(4)} (y'^2+1)^2)x^3+(2(y'^2+1))(-\frac{15}{2})y'y''^2+y'''(y'^2+1))x^2
 -27(y'^2+1)^2y''(y'^2+\frac{4}{3})x+33(y'^2+1)^3(y'^2+\frac{12}{11})y')$
\end{center}

(3, 3):
$X_{1} = y\p_{x} - x\p_{y}$, $X_{2} = (x^2-y^2+1)\p_{x} + 2xy\p_{y}$, $X_{3} = 2xy\p_{x} + (-x^2+y^2+1)\p_{y}$\\

The second and third order invariants are:
\begin{center}
$w_{2}=\dfrac{y''(1+x^{2}+y^{2})}{(1+y'^{2})^{\frac{3}{2}}}+\dfrac{2(y-xy')}{(1+y'^{2})^{\frac{1}{2}}}$,
$w_{3} = \dfrac{y'''(1+x^{2}+y^{2})^{2}}{(1+y'^{2})^{2}}-\dfrac{3y'y''^{2}(1+x^{2}+y^{2})^{2}}{(1+y'^{2})^{3}}$
\end{center}
 The fourth order invariant is then:
 \begin{center}
 $w_{4} = \lambda D_{x}(w_{3})$  for $\lambda = (1+x^{2}+y^{2})(1+y'^{2})^{\frac{-1}{2}}$
 \end{center}

The fourth order invariant:
\begin{center}
$\dfrac{1}{(y'^2+1)^{\frac{9}{2}}}((15y'^2y''^3-10y'y'''(y'^2+1)y''
 +y^{(4)}(y'^2+1)^2)x^6+(4(-\frac{15}{2}y'y''^2+(y'^2+1)y'''))(y'^2+1)x^5
 +(45y'^2(y^2+1)y''^3+(-12y'^4+6y'^2+18)yy''^2-30y'((y^2+1)y'''
 -\frac{6}{5}y'^3-\frac{6}{5}y')(y'^2+1)y''+(3*(\frac{4}{3}yy'y'''
 +y^{(4)}(y^2+1)))(y'^2+1)^2)x^4+(8(y'^2+1))(-\frac{15}{2}y'(y^2+1)y''^2
 -9y'y(y'^2+1)y''+((y^2+1)y'''-3y'^5-3y'^3)(y'^2+1))x^3
 +(45y'^2(y^2+1)^2y''^3-(24(y'^2-\frac{3}{2}))(y^2+1)y(y'^2+1)y''^2
 -(30(y'(y^2+1)^2y'''-(\frac{6}{5}((y'^2+1)y^2+y'^2))(y'^2+1)))(y'^2+1)y''
 +(3(\frac{8}{3}y'y(y^2+1)y'''+y^{(4)}y^4+2y^{(4)}y^2+(24y'^4+24y'^2)y
 +y^{(4)}))(y'^2+1)^2)x^2+(4(-\frac{15}{2}y'(y^2+1)^2y''^2-18y'y(y'^2+1)(y^2+1)y''
 +((y^2+1)^2y'''-18y^2y'(y'^2+1))(y'^2+1)))(y'^2+1)x+15y'^2(y^2+1)^3y''^3
 -(12(y'^2-\frac{3}{2}))(y^2+1)^2y(y'^2+1)y''^2-(10(y'(y^2+1)^2y'''
 -(\frac{18}{5}(y'^2+1))y^2))(y^2+1)(y'^2+1)y''+(4y'y(y^2+1)^2y'''
 +y^{(4)}y^6+3y^{(4)}y^4+(24y'^2+24)y^3+3y^{(4)}y^2+y^{(4)})(y'^2+1)^2)$
\end{center}

(11,3):
$X_{1} = \partial_{x}$, $X_{2} = x \partial_{x}$, $X_{3} = x^{2} \partial_{x}$

A zero and third order invariants are
\begin{center}
$w_0=y$, $w_3=\frac23 y'^{-3}y'''-y'^{-4}y''^2$
\end{center}

A direct computation using ${\cal D}=y'^{-1}D_x$ is simple in this case and we get a fourth order invariant:
\begin{center}
$w_4=y^{(4)}y'^{-4} + 6 y''^{3}y'^{-6} - 6 y'' y'''y'^{-5}$.
\end{center}

(17, 3):
$X_{1} = \p_{y}$, $X_{2} = x\p_{x} + y\p_{y}$, $X_{3} = 2xy\p_{x} + (x^2+y^2)\p_{y}$\\

The second and third order invariants are:
\begin{center}
$w_{2} = \dfrac{xy''-y'+y'^{3}}{(1-y'^{2})^{\frac{3}{2}}}$, $w_{3} = \dfrac{x^{2}y'''}{(y'^{2}-1)^{2}}- \dfrac{3x^{2}y'y''^{2}}{(y'^{2}-1)^{3}}$
\end{center}
and the fourth order invariant is:
\begin{center}
$w_{4} = \lambda D_{x}(w_{3})$ where $\lambda = x(1-y'^{2})^{\frac{-1}{2}}$
\end{center}

\begin{center}
$\dfrac{1}{((y'+1)^{\frac{9}{2}}(y'-1)^{\frac{9}{2}})}((15y'^2y''^3
+(-10y'^3y'''+10y'y''')y''+y^{(4)}(y'-1)^2(y'+1)^2)x^3+(2(y'-1))(y'^2y'''
-\frac{15}{2}y'y''^2-y''')(y'+1)x^2+(27(y'^2
-\frac{4}{3}))(y'-1)^2(y'+1)^2y''x+33y'^9-135y'^7+207y'^5-141y'^3+36y')$
\end{center}

\subsection*{B. The fundamental invariants of some algebras via Maple}

Here we present the fundamental differential invariants of the algebras (7, 6), (8, 8) and (16, 6). We have computed them using Maple 18 as doing them by hand is a tedious job. \\

$(7, 6)$: $X_{1} = \partial_{x}$, $X_{2} = \partial_{y}$, $X_{3} = x\partial_{x} + y \partial_{y}$, $X_{4} = y\partial_{x} - x\partial_{y}$, $X_{5} = (x^{2}-y^{2})\p_{x}+2xy \p_{y}$, $X_{6} = 2xy \p_{x} + (y^{2} - x^{2})\p_{y}$.\\

Fundamental Invariants:
\begin{center}
$\phi_{1} =
(1/4)\dfrac{1}{(y'^2y'''-3y'y''^2+y''')^3}((135y'^4-270y'^2-45)y''^6+(-180y'^5+180y')y'''y''^4
+30y^{(4)}(y'^2+1)^3y''^3-12(y'^2+1)^2((-\frac{10}{3}y'^2+\frac{20}{3})y'''^2+y'y^{(5)}(y'^2+1))y''^2
+40y'y'''y^{(4)}(y'^2+1)^3y''+4(y'^2+1)^3(-10y'y'''^3
+y^{(5)}(y'^2+1)y'''-\frac{5}{4})(y^{(4)})^2(y'^2+1)))$.
\end{center}

\begin{center}
$\phi_{2} =
\dfrac{1}{(\sqrt{-y'^2y'''+3y'y''^2-y'''}(y'^2y'''
 -3y'y''^2+y''')^4)}(y'^2+1)^3(-\frac{405}{4})*y''^9+\frac{405}{2}y'''y'y''^7
 -\frac{135}{4}y^{(4)}(y'^2-3)y''^6+((-\frac{225}{2}y'^2-\frac{405}{2})y'''^2
 +\frac{27}{2}y'y^{(5)}(y'^2-3))y''^5+(9(-\frac{5}{2}y'''(y'^2-7)y^{(4)}
 +y^{(6)}y'(y'^2+1)))y'y''^4+(-\frac{135}{4}(y'^2+1)^2(y^{(4)})^2
 -\frac{153}{2}y'''((-\frac{40}{51}y'^3+\frac{40}{51}y')y'''^2
 +(y'^2+1)y^{(5)}(y'^2-\frac{3}{17})))y''^3
 -(6(y'^2+1))(((-\frac{175}{4}y'^2-\frac{35}{4})y'''^2
 -\frac{9}{4}y'y^{(5)}(y'^2+1))y^{(4)}
 +y'y'''y^{(6)}(y'^2+1))y''^2
 +(24(y'^2+1))y'''((-\frac{15}{8}y'^3-\frac{15}{8}y')(y^{(4)})^2
 +((-\frac{85}{12}y'^2-\frac{25}{12})y'''^2
 +y'y^{(5)}(y'^2+1))y''')y''+(y'^2+1)^2((\frac{15}{4}y'^2
 +\frac{15}{4})(y^{(4)})^3-(\frac{9}{2}(-\frac{20}{9}y'y'''^2
 +y^{(5)}(y'^2+1)))y'''y^{(4)}
 +y'''^2y^{(6)}(y'^2+1)))$.
 \end{center}

(16, 6): $X_{1} = \p_{x}$, $X_{2} = \p_{y} $, $X_{3} = x\p_{x}$, $X_{4} = y\p_{y}$, $X_{5} = x^{2}\p_{x}$, $X_{6} = y^{2}\p_{y}$.\\

Fundamental Invariants:
\begin{center}
$\phi_{1} =
(1/6)d\frac{y'^{3}(12y'y'''y^{(5)}-15y'(y^{(4)})^{2}-18y''^{2}y^{(5)}
+60y''y'''y^{(4)}-40y'''^{3})}{(2y'y'''-3y''^{2})^{3}}$.
\end{center}

\begin{center}
$\phi_{2} =
\dfrac{1}{(\sqrt{-2y'y'''+3y''^{2}}(2y'y'''
-3y''^{2})^4)}(4((y'''^{2}y^{(6)}-\frac{9}{2}y'''y^{(4)}y^{(5)}
+\frac{15}{4}(y^{(4)})^{3})y'^{3}+(5y'''^{3}y^{(4)}+12y''y'''^{2}y^{(5)}
+(-3y''^{2}y^{(6)}-\frac{45}{2}y''(y^{(4)})^{2})y'''
+\frac{27}{4}y''^{2}y^{(4)}y^{(5)})y'^{2}+\frac{9}{4}y''(y''^{3}y^{(6)}
-10y''^{2}y'''y^{(5)}+\frac{70}{3}y''y'''^{2}y^{(4)}
-\frac{40}{3}y'''^{4})y'+\frac{27}{4}y''^{5}y^{(5)}
-\frac{45}{2}y''^{4}y'''y^{(4)}
+15y''^{3}y'''^{3})y'^{3})$.
\end{center}

\newpage
(8, 8): $X_{1} = \p_{x} $, $X_{2} = \p_{y}$, $X_{3} = x\p_{x} $, $X_{4} = y\p_{x}$, $X_{5} = x\p_{y} $, $X_{6} = y\p_{y}$, $X_{7} = x^{2}\p_{x} + xy\p_{y} $, $X_{8} = xy\p_{x} + y^{2} \p_{y}$.\\

Fundamental Invariants:
\begin{center}
$\phi_{1} =
(1/18)\dfrac{1}{(9y''^{2}y^{(5)}-45y''y'''y^{(4)}+40y'''^{3})^{8/3}}(162y''^{6}y^{(5)}y^{(7)}-189y''^{6}(y^{(6)})^{2}-810y''{^5}y'''y^{(4)}y^{(7)}+1134y''^{5}y'''y^{(5)}y^{(6)}
+1890y''^{5}(y^{(4)})^{2}y^{(6)}-2835y''^{5}y^{(4)}(y^{(5)})^{2}+720y''^{4}y'''^{3}y^{(7)}-3150y''^{4}y'''^{2}y^{(4)}y^{(6)}
-756y''^{4}y'''^{2}(y^{(5)})^{2}+13230y''^{4}y'''(y^{(4)})^{2}y^{(5)}-4725y''^{4}(y^{(4)})^{4}-12600y''^{3}y'''^{3}y^{(4)}y^{(5)}
-7875y''^{3}y'''^{2}(y^{(4)})^{3}+6720y''^{2}y'''^{5}y^{(5)}+31500y''^{2}y'''^{4}(y^{(4)})^{2}
-33600y''y'''^{6}y^{(4)}+11200y'''^{8})$.
\end{center}

\begin{center}
$\phi_{2} =
  (1/81)\dfrac{1}{(9y''^{2}y^{(5)}-45y''y'''y^{(4)}+40y'''^{3})^{4}}(-2551500y''^{6}(y^{(4)})^{6}
+7016625y''^{5}y'''^{2}(y^{(4)})^{5}
+(1530900(y''^{3}y^{(6)}+\frac{7}{2}y''^{2}y'''y^{(5)}
+45y'''^{4}))y''^{4}(y^{(4)})^4+(-656100y''^{7}y'''y^{(7)}-1760535y''^{7}(y^{(5)})^{2}
-5868450y''^{6}y'''^{2}y^{(6)}-73823400y''^{5}y'''^{3}y^{(5)}
-255150000y''^{3}y'''^{6})(y^{(4)})^{3}+(131220(161y''^{4}y'''^{2}(y^{(5)})^{2}+y''^{2}(y''^{4}y^{(7)}
+\frac{14}{3}y''^{3}y'''y^{(6)}+\frac{3640}{3}y'''^{5})y^{(5)}-\frac{7}{3}y''^{6}(y^{(6)})^{2}
+\frac{5}{4}y''^{5}y'''^{2}y^{(8)}+\frac{160}{9}y''^{4}y'''^{3}y^{(7)}+\frac{1540}{27}y''^{3}y'''^{4}y^{(6)}
+\frac{70000}{27}y'''^{8}))y''^{2}(y^{(4)})^{2}-(65610(\frac{168}{5}y''^{6}y'''(y^{(5)})^{3}
+(-\frac{21}{5}y''^{7}y^{(6)}+\frac{3640}{9}y'''^{4}y''^{4})(y^{(5)})^{2}+y''^{2}y'''(y''^{5}y^{(8)}
+\frac{40}{3}y''^{4}y'''y^{(7)}+\frac{280}{9}y''^{3}*y'''^{2}y^{(6)}+\frac{56000}{27}y'''^{6})y^{(5)}
-2y'''(y''^{7}y^{(6)}y^{(7)}+\frac{35}{9}y''^{6}y'''(y^{(6)})^{2}-\frac{20}{9}y''^{5}y'''^{3}y^{(8)}
-\frac{440}{27}y''^{4}y'''^{4}y^{(7)}-\frac{1400}{81}y''^{3}y'''^{5}y^{(6)}
-\frac{1120000}{729}y'''^{9})))y''y^{(4)}+653184y''^{6}y'''^{3}(y^{(5)})^{3}+6561y''^{4}(y''^{5}y^{(8)}
+16y''^{4}y'''y^{(7)}+\frac{224}{3}y''^{3}y'''^{2}y^{(6)}+\frac{150080}{81}y'''^{6})(y^{(5)})^{2}
-26244y''^{2}(y''^{7}y^{(6)}y^{(7)}+7y''^{6}y'''(y^{(6)})^{2}-\frac{20}{9}y''^{5}y'''^{3}y^{(8)}
-\frac{200}{9}y''^{4}y'''^{4}y^{(7)}+\frac{280}{27}y''^{3}y'''^{5}y^{(6)}
-\frac{1120000}{729}y'''^{9})y^{(5)}+20412y''^{9}(y^{(6)})^{3}-116640y''^{7}y'''^{3}y^{(6)}y^{(7)}
+129600y''^{5}y'''^{6}y^{(8)}+518400y''^{4}y'''^{7}y^{(7)}+44800000y'''^{12})$.
\end{center}

\end{section}

{\bf Acknowledgments}. FM thanks the NRF and Wits for research support.

\bibliographystyle{apa}

\end{document}